\documentstyle[12pt]{article}

\newcommand{\be}{\begin{equation}}
\newcommand{\ee}{\end{equation}}  
\newcommand{\eins}{\mbox{$\rule{2.5mm}{0.1mm} 
                          {\hspace{-2.7mm}1} 
                          {\hspace{-0.2mm}\rule{0.07mm}{2.7mm}}$}}

\newcommand{\vslash}{\mbox{$\not{\hspace{-0.8mm}v}$}}          % vslash    

\begin{document}
\bibliographystyle{references}
\baselineskip 18pt
\pagenumbering{arabic}          
\begin{titlepage} 
%\begin{flushright}
% IC/96/35\\
% MZ-TH/96-10 
% \end{flushright}

\begin{center} 

  {\large \bf Electromagnetic Transitions of Heavy Baryons in the 
$SU(2N_{f}) \otimes O(3)$ Symmetry}    

 \vspace*{1cm}  
 
 {\large {\rm Salam Tawfiq $^{\dagger}$, J.G. K\"{o}rner $^{\ddagger}$ and
Patrick J. O'Donnell $^{\dagger}$ }}
  
\vspace*{0.4cm}

  {\it $^{\dagger}$  Department of Physics, University of Toronto \\
  60 St. George Street, Toronto Ontario, M5S 1A7, Canada }
 \vspace*{0.5cm}

 {\it $^{\ddagger}$  Institut f\"{u}r Physik, Johannes
Gutenberg-Universit\"{a}t\\  
 Staudinger Weg 7, D-55099 Mainz, Germany}\\
 \vspace*{0.4cm}

 \vspace*{0.8cm}

 {\it May 1999} 
 
\vspace*{2cm}
       
\end{center} 
  
\begin {abstract}
Radiative decays of heavy baryons are analyzed within the Heavy Quark
Symmetry (HQS). It is shown that employing the light-diquark symmetries, 
the number of electromagnetic couplings among S-wave and P-wave states as
well as those between P-wave to S-wave transitions can be reduced
significantly. Using this constituent quark model picture the 
phenomenological implications of some of these decay modes are, also, 
discussed. 
\end{abstract}
  
\end{titlepage} 

%\newpage
\section{Introduction}
The strong decays are expected to dominate the branching rates of 
charmed baryons. In fact, most of the experimentally discovered channels of 
ground state charmed baryons are the one- and two- pion transitions \cite{exp}. 
In a recent paper, CLEO \cite{CLEO-new} presents the first observation of
the two narrow resonances decaying to $\Xi_c^+ \gamma$ and $\Xi_c^0 \gamma$. 
The mass differences $M(\Xi_c^+ \gamma)-M(\Xi_c^+)$ and 
$M(\Xi_c^0 \gamma)-M(\Xi_c^0)$ are measured to be 
($107.8\pm 3.2 $) ${\; \rm MeV}/c^2$ and ($107.0\pm 3.9 $) ${\; \rm MeV}/c^2$
respectively. These two states are identified as 
$\Xi_c^{+\prime}$ and $\Xi_c^{+\prime}$ which are the spin-$\frac{1}{2}$ 
partners of the already observed $\Xi_c^{*}$ states with spin-$\frac{3}{2}$.  
Since the masses of 
$\Xi_c^{'}$ states are below threshold for the decay to 
$\Xi_c \pi$, they will decay electromagnetically. Similar observations are 
expected also for some of the orbitally excited resonances. 

The recent observations 
of CLEO confirm the importance of the radiative channels even
though their contributions to some heavy baryon widths are relatively
small compared to other decay modes. The other important fact is that,  
the electromagnetic decay rates of some of the heavy
baryons can be used to determine some other couplings. 
The one-loop branching fraction $Br(\Xi_c^{0*}\rightarrow\Xi_c^0
\gamma)=\Gamma_{\gamma}/\Gamma_{\pi}$, for 
instance, is directly related \cite{a-etal,lsw} to the strong coupling
$g_{\bf 66\pi}$ or $g2$ as it is usually called in 
Heavy Hadron Chiral Perturbation Theory (HHCPT). Hence an
accurate 
measurement of $\Xi_c^{0*}$ electromagnetic decay fraction will allow the 
prediction of $g2$, which enters in many heavy baryon loop 
calculations.  
Although the electromagnetic strength is weaker than that of
the strong interaction, radiative channels are not phase space suppressed as
in the case of pion transitions. Therefore,
some radiative decay modes are expected to contribute significantly to
some heavy baryon branching fractions.

Radiative decays among ground states of heavy hadrons 
within the framework of HHCPT were
%\cite{g-states,cheng}% 
presented in \cite{a-etal,lsw,g-states,cheng}. It has been shown that
electromagnetic
interactions of heavy mesons contains only one coupling constant while  
heavy baryons transitions are determined in terms of two independent coupling
constants. Moreover, based on the same theory, a small enhancement in the
$\Sigma_Q^{*}\rightarrow\Lambda_Q\;\gamma$ radiative 
decay amplitudes was predicted \cite{savage} due to an electric quadrapole
$E2$ interactions.  
On the other hand, electromagnetic transitions 
of excited $\Lambda_Q$ baryons
were analyzed in the (HHCPT) \cite{cho} and 
in the bound state picture \cite{chow}. They show that the charmed baryons 
decay channels are severely suppressed, however, the bottom baryons may
have 
significant branching ratios. Some interesting predictions for charmed
baryons within the frame of a relativistic quark model (RTQM) are
presented in ref. \cite{iklr}. 
Finally, the light-cone QCD sum rules in the
leading order of HQET were also used \cite{zd} to analyze radiative decays
of heavy baryons.   

In a number of recent papers \cite{pion,hklt} it was shown that, more
phenomenological predictions can be achieved if Heavy Quark Symmetry (HQS) 
is supplemented by the light diquark $SU(2N_f)\times O(3)$ symmetry, with
$N_f$ being the number of light flavors.
In fact, the large number of HQS couplings, both the weak and strong
couplings, are reduced to few effective quark couplings.   
In this work we shall follow the same constituent quark model 
picture to describe radiative transitions among heavy baryon states. This
becomes simple and justified since 
the radiative decay mechanism is quite similar, for instance, to that of
single-pion transitions employed in \cite{pion}. 
The rest of the paper is
organized as follows: In the next section, we provide the HQS predictions for 
heavy baryon ground states and P-wave to S-wave electromagnetic
transitions. 
In section 3, we show how the light-diquark
symmetries can be used to relate the various HQS electromagnetic coupling to 
a few constituent quark model effective couplings. 
The decay rates of some charmed and bottom baryons radiative
modes are
predicted in Sec.4. 
We summarize our conclusions in Sec.5, while in the Appendix we give the
detailed analysis
for electromagnetic transitions among P-wave states.       
%=========================================-------------------
\section{HQS Matrix Elements}
To leading order in the heavy quark limit, the heavy baryon spin wave
function can be written in the simple covariant and general form 
\be\label{Psi}
\Psi^{B_Q}_{\alpha\beta\gamma}(v,q)=(\phi_{\mu_{1}
\dots\mu_{j}}(v,q))_{\alpha\beta}(\psi^{\mu_{1}
\dots\mu_{j}}(v))_{\gamma}\; .
\ee  
Here, $j$ refer to the spin of the diquark system, $v$ is the velocity of
the heavy baryon and $q$ represents the photon momentum. 
In Eq. (\ref{Psi}), we have kept the Dirac indices
$\alpha$, $\beta$ and $\gamma$, while the flavor
and color indices, which can easily be included, are omitted.
Details of the formalism to construct these functions, for arbitrary 
orbital angular momentum, can be found in \cite{htk,kkp}.
The diquark spin wave functions
$(\phi_{\mu_{1}\dots\mu_{j}}(v))_{\alpha\beta}$
for S-wave and P-wave states of heavy baryons 
are listed in table \ref{swf}. In this table, we distinguish
between two kind of excitations, namely the K- and k-excitations which,
respectively, 
correspond to {\it symmetric} and {\it antisymmetric} states when 
interchanging the light diquark indices. These, actually, can be identified
as the canonical momenta of the well know $\lambda$- and $\rho$-
coordinates successfully used to determine the w
charmed baryon spectrum
\cite{isgur-karl}. 

The heavy quark spin wave functions $(\psi^{\mu_{1}
\dots\mu_{j}}(v))_{\gamma}$ in Eq. (\ref{Psi}) can be interpreted as a
"superfield" corresponding to the two heavy quark symmetry degenerate 
states with spin $j-1/2$ and $j+1/2$. It is 
generally written in terms of the Dirac spinor $u$ and the
Rarita-Schwinger 
spinor $u_{\mu}$. The spin-$\frac{1}{2}$ ground state $\Lambda_{Q}$ 
field is simply given by 
\be
(\psi)_{\gamma}=u_{\gamma}, 
\ee
a similar form can be used to describe the heavy quark field of P-wave
state $\Sigma_{Q0}$. 
The $\Sigma_{Q}$, which are spin-$1/2$
and $3/2$ degenerate are described by the superfield 
\be\label{spin-1}
(\psi_{\mu})_{\gamma}=\left\{\begin{array}{r}
\frac{1}{\sqrt{3}}\gamma^{\perp}_{\mu} \gamma_{5}u \\
u_{\mu} \end{array}
\right\}_{\gamma} \,
\ee
The two P-wave degenerate states $\Lambda_{Q1}$ and
$\Sigma_{Q1}$ heavy quark superfields are also given by Eq.
(\ref{spin-1}).
Finally, the $\Sigma_{Q2}$ superfield can be written as 
\be
(\psi_{\mu1\mu2})_{\gamma}=\left\{\begin{array}{r}
\frac{1}{\sqrt{10}}\{\gamma^{\perp}_{\mu1} \gamma_{5}u_{\mu2}\}_0 \\
u_{\mu1\mu2} \end{array} 
\right\}_{\gamma} \; .
\ee
Here, $\{\gamma^{\perp}_{\mu1} \gamma_{5}u_{\mu2}\}_0$
represents a traceless symmetric tensor.
%****************************Table -1*********************************     
\begin{table}
\caption{\label{swf}S-wave  and P-wave heavy baryons covariant diquark
spin wave functions. Here, we have $\chi^0=(\vslash
+1)\gamma_5C {\rm \; and \; }
\chi^{1,\mu}=(\vslash+1)\gamma_{\perp}^{\mu}C$, with $C=i\gamma_0\gamma_2$ and
$\gamma_{\perp \mu}=\gamma_{\mu}-\vslash v_{\mu}$.}
\vspace{5mm} 
\renewcommand{\baselinestretch}{1.2}
\small \normalsize
\begin{center}
\begin{tabular}{|c|c|c|}
\hline \hline
& $j^{P}$ &  ${\phi}_{\alpha\beta} $ \\ 
\hline \hline
\multicolumn{3}{|l|} {  S-wave states } \\
 $\Lambda_{Q}$ & $ 0^{+}$  & $(\chi^{0})_{\alpha\beta}$ \\
\hline
$ \Sigma_{Q} $ & $1^{+}$ &
         $(\chi^{1,\mu })_{\alpha\beta}  $ \\ 
\hline\hline
\multicolumn{3}{|l|}{ Symmetric P-wave states}  \\
$ \Lambda_{Q1} $ & $1^{-}$  & 
     $ (\chi^{0} K_{\perp}^{\mu})_{\alpha\beta} $ \\ 
\hline
$\Sigma_{Q0} $ & $0^{-}$  & $\frac{1}{\sqrt{3}} (\chi^{1,\mu}
 K_{\perp \mu} )_{\alpha\beta} $\\ 
\hline
$ \Sigma_{Q1} $ & $1^{-}$  & $
 \frac{i}{\sqrt 2} ( \varepsilon_{\mu\nu\rho\delta}  \chi^{1,\nu} 
  K_{\perp}^{\rho}v^{\delta} )_{\alpha\beta} $ \\ 
\hline
$ \Sigma_{Q2} $& $2^{-}$ &
   $ \frac{1}{2}(\{ \chi^{1,\mu_1}K_{\perp}^{\mu_{2}}\}_{0})_{\alpha\beta}$ \\ 
\hline \hline
\multicolumn{3}{|l|}{ Antisymmetric P-wave states } \\
$ \Sigma_{Q1} $ & $1^{-}$  & 
     $ (\chi^{0} k_{\perp}^{\mu})_{\alpha\beta} $ \\ 
\hline
$\Lambda_{Q0} $ & $0^{-}$ & $\frac{1}{\sqrt{3}} (\chi^{1,\mu}  
 k_{\perp\mu} )_{\alpha\beta}  $ \\ 
\hline
$ \Lambda_{Q1} $ & $1^{-}$  & $ 
 \frac{i}{\sqrt 2} ( \varepsilon_{\mu\nu\rho\delta} \chi^{1,\nu} 
  k_{\perp}^{\rho}v^{\delta} )_{\alpha\beta} $ \\ 
\hline
$ \Lambda_{Q2} $& $2^{-}$ &
   $\frac{1}{2} (\{\chi^{1,\mu_1} k_{\perp}^{\mu_{2}}\}_{0})_{\alpha\beta}$ \\ 
\hline \hline
\end{tabular}
\renewcommand{\baselinestretch}{1}
\small \normalsize
\end{center}
\end{table}                                                                        

To leading order in the limit $m_Q\rightarrow \infty$, HQS requires the
photon to couple to the light diquark only, while the heavy quark
propagates without being affected by the photon emission.
Therefore, electromagnetic decays proceed through interactions that
change the spin of the light degrees of freedom. Then,
as a consequence of HQS, the photon orbital momentum
relative to the diquark $l_{\gamma}$ is identical to that relative to the 
baryon $L_{\gamma}$. Therefore,            
radiative transition amplitudes between heavy baryons take the form  
\begin{eqnarray}
M^{\gamma}&=& \langle \gamma (q), B_{Q2}(v) \mid T \mid B_{Q1}(v) \rangle
\nonumber\\[2mm]
&=&{\cal H}^{\mu_1 \cdots \mu_{j_1}}_{\nu_1 \cdots \nu_{j_2}} 
{\cal M}^{\nu_1 \cdots \nu_{j_2}}_{\mu_1 \cdots \mu_{j_1}} \; ,
\end{eqnarray} \label{trans}   
where ${\cal H}(v)$ and ${\cal M}(v,q)$, respectively, represents the 
heavy quark and the diquark transition matrix elements and $j_1$ and $j_2$ 
are the diquark spin degrees of freedom in the initial and final heavy 
baryons. In the heavy quark limit, ${\cal} H(v)$ is a function of the heavy 
quark velocity which is identical to the baryon velocity $v$ and can be 
constructed in terms of the superfields defined earlier.  
 The tensors ${\cal M}(v,q)$, which are also a function of the 
 photon momentum $q$, are constructed to ensure HQS, parity and gauge 
 invariance. For simplicity and to display the similarity with one-pion
transitions, they 
 can be written in terms of multipole amplitudes with multipolarity 
 $2^{J_{\gamma}}$ which correspond to a 
 specific photon angular momentum $J_{\gamma}$. Their 
contributions to the decay rates are proportional to 
${\mid \vec{q}\mid}^{2J_{\gamma}+1}$, with $\vec{q}$ being the photon 
momentum in the centre of mass of the decaying baryon. 

As can be seen from Table (\ref{tab-HQS-a}), HQS predicts that S-wave to
S-wave
heavy baryon electromagnetic transitions are described in terms of {\bf
three} independent couplings $g_1$, $g_2$ and $f_2$. And there are {\bf
eleven} such couplings for 
transition from each of the symmetric and antisymmetric P-wave  
to S-wave states. The HQS matrix elements of
these transitions are summarized in Tables \ref{tab-HQS-a} and
 \ref{tab-HQS-b}.
Furthermore, as will be discussed in the Appendix, each of the 
P-wave to P-wave transitions, diagonal ($K\rightarrow K$) or 
($k \rightarrow k$) and off diagonal ($k \rightarrow K$), 
are determined in terms of {\bf fifteen} couplings. Their 
HQS matrix elements are given in Tables \ref{tab-HQS-b}.
%here ($A:=(\alpha;a;i)$) with $\alpha$ being a Dirac index, $a$ is
%a flavor index and $i$ refer to the color index. The diquark constituent wave
%function $D_{AB}$ is completely symmetric in the light quark degrees of 
%freedom.        
 
%****************** T A B L E  2-a *****************************************
\begin{table}
\caption[dumy16]{\label{tab-HQS-a} Electromagnetic transition amplitudes 
for S-wave to S-wave and symmetric P-wave to S-wave heavy baryon states 
as predicted by HQS. Here, photons are in definite 
electric ($EJ_{\gamma}$) or magnetic ($MJ_{\gamma}$) multipoles state. Similar 
amplitudes can be written for antisymmetric P-wave to S-wave heavy baryons 
with new couplings $f_i^{(k)}$.}
%\begin{center}   
\vspace{-2mm}
\renewcommand{\baselinestretch}{1.2}
\small \normalsize
\setlength{\leftmargin}{-1.0cm}
\begin{center}
%\footnotesize {  
\small {    
\begin{tabular}{|c||c|c|c|}
\hline \hline
 $B_{Q} \rightarrow B^{\prime}_Q + \gamma$     &
 Multipole state     &
 ${\cal H}^{\mu_1 \cdots\mu_{j_1}}_{\nu_1\cdots \nu_{j_2}}(v) $   &
 ${\cal M}^{\nu_1 \cdots\nu_{j_2}}_{\mu_1\cdots \mu_{j_1}}(v,q) $\\
\hline \hline      
 $\Sigma_{Q}\rightarrow \Lambda_Q$ & $M1$ &
 ${\bar \psi}(v)\psi^{\mu}(v)$     &     
$ig_1{\tilde F}^{\alpha \beta}g_{\mu\alpha}v_{\beta}$\\     
\hline
 $\Sigma_{Q} \rightarrow \Sigma_Q$ & $ M1$,$E2$ & 
  ${\bar \psi}^{\nu}(v)\psi^{\mu}(v) $  &
 $ \begin{array}{c}
 g_2 F_{\alpha \beta}g_{\mu}^{\alpha}g_{\nu}^{\beta} +\\
  f_2 F_{\alpha\beta}(2q_{\mu}g_{\nu}^{\alpha}
  v^{\beta}+v \hspace{-0.7mm}\cdot\hspace{-0.7mm} q g_{\mu}^{\alpha}
 g_{\nu}^{\beta}) \end {array}   $\\         
\hline \hline  
%\normalsize 
\multicolumn{1}{l}{$B_{Qi} \rightarrow B^{\prime}_Q +\gamma$} &
\multicolumn{2}{c}{K-multiplet}\\
\hline\hline  
%\hspace*{-0.8cm} 
$\Lambda_{Q1}\rightarrow \Lambda_Q$ & $ E1$ &
 ${\bar \psi}(v)\psi^{\mu}(v) $ &
 $f_1^{(K)}F_{\alpha \beta} g_{\mu}^{\alpha} v^{\beta}$\\     
\hline              
$\Lambda_{Q1}\rightarrow \Sigma_Q$ & $ E1$,$M2$ &  
  ${\bar \psi}^{\nu}(v)\psi^{\mu}(v) $ &
 $\begin{array}{c} 
if_2^{(K)}{\tilde F}^{\alpha \beta}g_{\mu\alpha} g_{\nu}^{\beta} +\\ 
  ig_2^{(K)}{\tilde F}^{\alpha \beta}(2q_{\mu}
  g_{\nu\alpha}v_{\beta}+v \hspace{-0.7mm}\cdot\hspace{-0.7mm}q
 g_{\mu\alpha} g_{\nu\beta}) \end{array}  $\\      
\hline              
%\hspace*{-0.8cm} 
$\Sigma_{Q0}\rightarrow \Sigma_Q$ & $ E1$ &  
  ${\bar \psi}^{\mu}(v)\psi(v) $ &
 $ f_3^{(K)} F_{\alpha \beta}g_{\mu}^{\alpha} v^{\beta} $\\      
\hline           
%\hspace*{-0.8cm} 
$\Sigma_{Q1}\rightarrow \Lambda_Q$ & $E1$ &  
 ${\bar \psi}(v)\psi^{\mu}(v) $ &
\hspace*{-2cm} $ f_4^{(K)}F_{\alpha \beta} g_{\mu}^{\alpha} v^{\beta}$\\     
\hline              
%\hspace*{-0.8cm} 
$\Sigma_{Q1} \rightarrow \Sigma_Q$ & $ E1$,$M2$ &  
 ${\bar \psi}^{\nu}(v)\psi^{\mu}(v) $ & 
 $ \begin{array}{c} 
 i f_5^{(K)}{\tilde F}^{\alpha \beta}g_{\mu\alpha} g_{\nu\beta}+ \\ 
 i g_5^{(K)}{\tilde F}^{\alpha \beta}(2q_{\mu}
  g_{\nu\alpha}v_{\beta}+v \hspace{-0.7mm}\cdot\hspace{-0.7mm}q
  g_{\mu\alpha} g_{\nu\beta}) 
\end{array} $\\   
\hline
 $ \Sigma_{Q2} \rightarrow\Lambda_Q$ & $M2$ &  
${\bar \psi}(v)\psi^{\mu_1\mu_2}(v) $ &   
\hspace*{-0.8cm} $ig_6^{(K)}{\tilde F}^{\alpha \beta} 
q_{\mu_1} g_{\mu_2\alpha} v_{\beta} $\\
\hline
 $\Sigma_{Q2}\rightarrow \Sigma_Q$ & 
$ \begin{array}{c}  E1,M2, \\  E3 \end{array} $ &  
${\bar \psi}^{\nu}(v)\psi^{\mu_1\mu_2}(v) $ &
 $  \begin{array}{c}  
f_7^{(K)}F_{\alpha \beta}g_{\mu_{1}}^{\alpha} g_{\mu_{2}\nu}v^{\beta}+ \\
 g_7^{(K)}F_{\alpha \beta} 
 ( 2q_{\mu_{2}} g_{\mu_{1}}^{\alpha}g_{\nu}^{\beta}+ 
v \! \cdot \! q g_{\mu_{2} \nu} g_{\mu_{1}}^{\alpha}v^{\beta})\\
 + f_8^{(K)}F_{\alpha \beta} 
 [q_{\mu_{2}} (q_{\nu}g_{\mu_{1}}^{\alpha}+q_{\mu_1}
 g_{\nu}^{\alpha}) v^{\beta}+ \\
(v \! \cdot \! q) \{ q_{\mu_2}g_{\mu_{1}}^{\alpha}g_{\mu}^{\beta}+
 (v \! \cdot \! q)g_{\mu_2 \nu}g_{\mu_1}^{\alpha}v^{\beta} \}]
\end{array} $\\
\hline \hline      
\end{tabular}  
}
\end{center}
\renewcommand{\baselinestretch}{1}
\small \normalsize
\end{table}      
%\newpage
%=============================================== 
\section{Constituent Quark Model Couplings}
As it was shown in the previous section, there are many HQS couplings
required to described S-wave and P-wave heavy
baryons radiative transitions. In fact, there are at least forty such
couplings which are summarized in Tables
\ref{tab-HQS-a} and \ref{tab-HQS-b}. Therefore, even to leading
order in heavy quark symmetry limit there are only few  phenomenological
predictions
that can be made regarding the decay rates of these modes. This represents
the main motivation to try and go beyond HQS by using symmetries of the
light-diquark system. We shall try to achieve that by   
employing a constituent quark model with an underlying 
$SU(2N_f)\otimes O(3)$ diquark symmetry. These extra symmetries relate
the large number of HQS couplings to just few effective constituent 
couplings. 

The diquark decay amplitudes for a
 $\{j_1\}\rightarrow \{j_2\}+\gamma$ transition can be written as
 \be\label{M-diquark-gen}
 {\cal M}^{\nu_1\nu_2 \cdots \nu_{j_2}}_{\mu_1\mu_2\cdots \mu_{j_1}}=
 \left( {\bar \phi}^{\nu_1 \nu_2 \cdots \nu_{j_2}} \right)^{A B}
\left({\cal O}\right)^
{A^{'} B^{'}}_{A B}
\left({\phi}_{\mu_1\mu_2 \cdots \mu_{j_1}}\right)_{A^{'} B^{'}} \; .  
\ee  
 Here ${\cal O}$ is an effective electromagnetic transition operator 
 and ($A\equiv (\alpha;a;i)$) $\alpha$ is Dirac index, $a$ is a
flavor 
 index and $i$ refers to the color index. Once again, flavor and color
indices will be omitted since they are irrelevant in our case and can
easily be included later on in the decay rate formulae.  
 When analyzing transitions among 
higher resonance states, it is more convenient to rewrite the diquark 
system spin wave functions $(\phi)_{\alpha\beta}$ of Table \ref{swf} such
that the angular momentum excitations $K$ and $k$ are 
factorized which are then absorbed in the transition effective operator. 
One then writes
\be
\left({\phi}_{\mu_1\mu_2 \cdots \mu_{j}}\right)_{\alpha\beta}=
\left({\phi}^{\lambda_1\lambda_2\cdots \lambda_{L}}_
{\mu_1\mu_2 \cdots \mu_{j}}\right)_{\alpha\beta}
q_{\lambda_1}q_{\lambda_2}\cdots q_{\lambda_{L}} \; ,
\ee
where each of the $q_{\lambda_i}$, which corresponds to either $K$ or $k$, 
represents one unit of the light diquark angular momentum excitation $L_i$.
Therefore, the matrix elements in Eq. (\ref{M-diquark-gen}) can be
rewritten in the general form  
\be\label{M-diquark}
 {\cal M}^{\nu_1\nu_2 \cdots \nu_{j_2}}_{\mu_1\mu_2\cdots \mu_{j_1}}=
 \left( {\bar \phi}^{\nu_1 \nu_2 \cdots \nu_{j_2}}_{\eta_1\eta_2\cdots 
 \eta_{L_2}} \right)^{\alpha\beta}
\left({\cal O}^{\eta_1\eta_2\cdots \eta_{L_2}}_
{\lambda_1\lambda_2\cdots \lambda_{L_1}}(q)\right)^
{\alpha^{'}\beta^{'}}_{\alpha\beta}
\left({\phi}^{\lambda_1\lambda_2\cdots \lambda_{L_1}}_
{\mu_1\mu_2 \cdots \mu_{j_1}}\right)_{\alpha^{'}\beta^{'}} \; ,  
\ee 
with $L_1$ and $L_2$ being the initial- and final-diquark 
states angular excitations, respectively. 

The explicit form of the effective operators 
${\cal O}^{\eta_1\eta_2\cdots \eta_{L_2}}_{\lambda_1\lambda_2\cdots 
\lambda_{L_1}}(q_{\perp})$ will mainly depend on the multipolarity involved
in the decay process. These N-body operators conserve parity and are
gauge invariant; they should also project out the appropriate partial 
wave amplitudes involved in the transition. In general,
${\cal O}^{\eta_1\eta_2\cdots \eta_{L_2}}_{\lambda_1\lambda_2\cdots 
\lambda_{L_1}}(q)$ are constructed from the Dirac matrices, the metric 
tensor $g_{\mu\nu}$, the 
baryon velocity $v_{\alpha}$, the photon momentum $q$, and the 
electromagnetic tensor $F_{\alpha\beta}$ or its dual 
${\tilde F}_{\alpha\beta}=
 \frac{1}{2}\epsilon_{\alpha\beta\delta\rho}F^{\delta\rho}$. 
 
 In the constituent quark model and as required by HQS, the 
photon is emitted by one of the light quarks forming the heavy baryon 
diquark state. Therefore, the total heavy baryon electromagnetic coupling 
is the sum of the photon coupling to each of the light constituent quarks. 
As in pion transitions, one expects that the leading contributions to 
the quark-pion effective operator 
${\cal O}^{\eta_1\eta_2\cdots \eta_{L_2}}_{\lambda_1\lambda_2\cdots 
\lambda_{L_1}}(q)$ are those from 
one-body operators. These assumptions are confirmed by the $1/N_c$ expansion 
which shows that two-body strong transition operators are non leading for
pion 
couplings to S-wave heavy baryons and can be neglected in the constituent
quark model 
approach \cite{cgkm,py-NC}. On the other hand, the situation for P-wave
states 
in the $1/N_c$ expansion is still not clear. An explicit numerical analysis of 
the P-wave to S-wave strong decays \cite{cgkm} showed that the coefficients 
of the two-body operators are small. However, in ref.\cite{py-NC} it
was shown that two-body operators for S-wave 
pion couplings to P-wave states are not suppressed with respect 
to one-body operators but are nevertheless proportional to them. Here, we
shall consider one-body 
operators only and leave the effect of two-body operators for a future
work. 

 Radiative transitions of S-wave to S-wave states are induced by magnetic
dipole $M1$ and electric quadrapole $E2$ interactions.     
Therefore, the appropriate and unique effective magnetic dipole operator, with
decay constant $g_{M1}$, that couples the photon to the 
constituent quarks can be written as
\be\label{M1s-s}
\left({\cal O}^{M_1}\right)^{\alpha\beta}_{\alpha^{'}\beta^{'}} =
i\frac{g_{M1}}{2} 
\left\{\mu_{q_1}(\sigma^{\delta\rho})^{\alpha}_{\alpha^{'}}
 \otimes (\eins)^{\beta}_{\beta^{'}}
          + \,\mu_{q_2}(\eins )^{\alpha}_{\alpha^{'}}\otimes 
(\sigma^{\delta\rho})
^{\beta}_{\beta^{'}}\right\}
 \, F_{\delta\rho}   \; ,
\ee                   
where $\mu_{qi}=e_{qi}\frac{e}{2m_q}$ is the magnetic moment of the
effective light 
quark. Since we are considering real photons, there is in fact no one-body
operator that can be constructed to describe
electric quadrapole $E2$ transitions.
Therefore, $E2$ radiative 
transitions, among ground state heavy baryons, are either forbidden or they 
are due to two-body interactions and will be suppressed. A two-body $E2$
operator of vector-type gives no contribution to the radiative decay 
matrix elements. However,
an axial-vector type interaction operator does contribute and has the form  
\[
\left({\cal O}^{E2}\right)^{\alpha\beta}_{\alpha^{'}\beta^{'}} =
\frac{f_{E2}}{2} 
\left\{\mu_{q_1}(\gamma^{\delta}\gamma_5)^{\alpha}_{\alpha^{'}}
 \otimes (\gamma^{\rho}\gamma_5)^{\beta}_{\beta^{'}}
          + \,\mu_{q_2}(\gamma^{\rho}\gamma_5 )^{\alpha}_{\alpha^{'}}\otimes 
(\gamma^{\delta}\gamma_5)^{\beta}_{\beta^{'}}\right\}
\] \nonumber \be\label{E2s-s} 
\left[q_{\rho} F_{\delta\eta}v^{\eta}+(v\cdot q)\; F_{\delta\rho} \right]
\ee                                             
Next, we consider radiative transitions from P-wave excited states to the 
ground S-wave states which correspond 
to $E1$, $M2$ and $E3$ transitions. The relevant effective one-body 
operators have the form
\be\label{E1-K(p-s)}
\left({\cal O}_{\lambda}^{E1}\right)^{\alpha\beta}_{\alpha^{'}\beta^{'}} =
i\frac{f_{E1}^{(K)}}{2} 
\left\{\mu_{q_1}(\sigma^{\delta\rho})^{\alpha}_{\alpha^{'}}
 \otimes (\eins)^{\beta}_{\beta^{'}}
          + \,\mu_{q_2}(\eins )^{\alpha}_{\alpha^{'}}\otimes 
(\sigma^{\delta\rho})
^{\beta}_{\beta^{'}}\right\}
 \, F_{\rho\eta}g_{\delta\lambda}v^{\eta} \;\;\; ,
\ee                 
\[
\left({\cal O}_{\lambda}^{M2}\right)^{\alpha\beta}_{\alpha^{'}\beta^{'}} =
i\frac{g_{M2}^{(K)}}{2} 
\left\{\mu_{q_1}(\sigma^{\delta\rho})^{\alpha}_{\alpha^{'}}
 \otimes (\eins)^{\beta}_{\beta^{'}}
          + \,\mu_{q_2}(\eins )^{\alpha}_{\alpha^{'}}\otimes 
(\sigma^{\delta\rho})
^{\beta}_{\beta^{'}}\right\}
\]  \nonumber
 \be\label{E2-K(p-s)}  
 \left[2q_{\rho} F_{\delta\lambda}+(v\cdot q)\;
F_{\rho\eta}g_{\delta\lambda}v^{\eta}\right] \; .
\ee    
\[
\left({\cal O}_{\lambda}^{E3}\right)^{\alpha\beta}_{\alpha^{'}\beta^{'}} =
i\frac{f_{E3}^{(K)}}{2} 
\left\{\mu_{q_1}(\sigma^{\delta\rho})^{\alpha}_{\alpha^{'}}
 \otimes (\eins)^{\beta}_{\beta^{'}}
          + \,\mu_{q_2}(\eins )^{\alpha}_{\alpha^{'}}\otimes 
(\sigma^{\delta\rho})
^{\beta}_{\beta^{'}}\right\}
\]  \nonumber
 \be\label{E3-K(p-s)}  
 \left[q_{\rho}(2q_{\lambda} F_{\delta}^{\eta}v^{\eta}+(v\cdot q)\;
F_{\rho\eta}g_{\delta\lambda}v^{\eta})+(v\cdot q)^2\;
F_{\rho\eta}g_{\delta\lambda}v^{\eta})\right]
\ee
Transitions involving $J_{\gamma}=3$, however, are expected to be
suppressed and give practically no contribution to the electromagnetic
decay rate.  
The appropriate effective couplings ${\cal O}_{\lambda\xi}$ for
electromagnetic transitions among P-wave states are presented in the
appendix.    
 
As we mentioned before, k-multiplet P-wave spin wave functions are
antisymmetric under the exchange of the light quark indices while those of 
the k-multiplet are symmetric. Therefore, it
is possible to use these symmetries in writing the appropriate k-multiplet
photon transition operators from those given
in Eqns. (\ref{E1-K(p-s)}-\ref{E3-K(p-s)}) for the K-multiplet.
 One simply replaces the plus sign, in the
curly parenthesis, by a minus sign and the coupling constants 
$f_{Ei}^{(K)}$ and $g_{Mi}^{(K)}$ by the new couplings $f_{Ei}^{(k)}$ and
$g_{Mi}^{(k)}$ in the corresponding operators
${\cal O}_{\lambda}^{E1}$, ${\cal O}_{\lambda}^{M2}$, 
${\cal O}_{\lambda}^{E3}$ and        
${\cal O}_{\lambda\zeta}^{M1}$ and ${\cal O}_{\lambda\zeta}^{E2}$.    

The last step is to evaluate the matrix elements in Eq. (\ref{M-diquark})
using the corresponding operators 
${\cal O}$, ${\cal O}_{\lambda}$ and ${\cal O}_{\lambda\zeta}$ along with 
the light-diquark spin wave functions of table \ref{swf}. 
After a little bit of algebra, the
radiative transition matrix elements for S-wave to S-wave, P-wave to S-wave and 
P-wave to P-wave heavy baryon states are
calculated. Neglecting flavor factors, one obtains
relations for the electromagnetic coupling constants among heavy baryon
states. Defining    
\begin{eqnarray} \label{delta+-}     
\mu^{(+)}= (\frac{e_{q_1}}{2m_{q_1}}+ \frac{e_{q_2}}{2m_{q_2}}) \nonumber\\ 
\mu^{(-)}= (\frac{e_{q_1}}{2m_{q_1}}- \frac{e_{q_2}}{2m_{q_2}}),
\end{eqnarray}     
with $e_{q_i}$ and $m_{q_i}$ being the charge and the mass of the light
quark, we obtain\\
 {\underline { \it\bf S-wave to S-wave:}   
\begin{eqnarray} \label{s-s}   
g_1=\mu^{(-)}g_{M1} \nonumber\\ 
g_2=\mu^{(+)}g_{M1}.
\end{eqnarray}    
{\underline {\it \bf P-wave (K-multiplet) to S-wave}:   
\[
f_1^{(K)}=0\;\;,\;\; f_2^{(K)}=\frac{1}{2}\mu^{(+)}f_{E1}^{(K)} \;\;,\;\; 
f_3^{(K)}=-\frac{1}{\sqrt{3}}\mu^{(+)}f_{E1}^{(K)} \;\;,\;\;
f_4^{(K)}=-\frac{1}{\sqrt{2}}\mu^{(-)}f_{E1}^{(K)} \;\;,\;\;   \]  \nonumber
\be\label{Kf-p-s}   
f_5^{(K)}=\frac{1}{2\sqrt{2}}\mu^{(+)}f_{E1}^{(K)} \;\;,\;\;
 f_7^{(K)}=-\frac{1}{2}\mu^{(+)}f_{E1}^{(K)} \;\;,\;\;   
 f_8^{(K)}=-\frac{1}{2}\mu^{(+)}f_{E3}^{(K)}  \;\;,\;\;                              
\ee
and
\[
g_2^{(K)}-\mu^{(+)}g_{M2}^{(K)}\;\;,\;\; 
g_5^{(K)}=\sqrt{2}\mu^{(+)}g_{M2}^{(K)} \;\;,\;\;
\] \nonumber
\be\label{Kg-p-s}  
g_6^{(K)}=\mu^{(-)}g_{M2}^{(K)} \;\;,\;\; 
g_7^{(K)}=-\mu^{(+)}g_{M2}^{(K)}  
\ee                                   
{\underline  {\it\bf P-wave to P-wave (K-multiplet) }:   
\[
F_1^{(K)}=0\;\;,\;\; 
F_3^{(K)}=-\frac{1}{\sqrt{2}}\mu^{(+)}F_{E2}^{(K)} \;\;,\;\;  
F_5^{(K)}=\frac{1}{4}\mu^{(+)}F_{E2}^{(K)} \;\;,\;\;  
F_6^{(K)}=-\frac{1}{2}\mu^{(+)}F_{E2}^{(K)} \;\;, 
 \]  \nonumber
\be\label{Kf-p-p}    
F_7^{(K)}=\frac{1}{2\sqrt{3}}\mu^{(+)}F_{E2}^{(K)} \;\;,\;\;
F_8^{(K)}=\frac{1}{2\sqrt{2}}\mu^{(+)}F_{E2}^{(K)} \;\;,\;\;  
F_9^{(K)}=-\frac{1}{2}\mu^{(+)}F_{E2}^{(K)}  
\ee  
and
\[
G_1^{(K)}=0\;\;,\;\; 
G_2^{(K)}=\frac{1}{\sqrt{3}}\mu^{(-)}G_{M1}^{(K)}\;\; ,\;\;  
G_3^{(K)}=\frac{1}{2\sqrt{2}}\mu^{(+)}G_{M1}^{(K)} \;\; ,\;\;
\]\nonumber 
\[
G_4^{(K)}=\sqrt{\frac{2}{3}}\mu^{(+)}G_{M1}^{(K)} \;\; ,\;\;
G_5^{(K)}=-\frac{1}{4}\mu^{(+)}G_{M1}^{(K)} \;\; , \;\;   
G_6^{(K)}=-\frac{1}{2}\mu^{(+)}G_{M1}^{(K)} \;\; , \;\; 
\]\nonumber 
\be\label{Kg-p-p}   
G_8^{(K)}=-\frac{1}{\sqrt{2}}\mu^{(+)}G_{M1}^{(K)} \;\; ,\;\;  
G_9^{(K)}=\frac{1}{2}\mu^{(+)}G_{M1}^{(K)}  
\ee
One can obtain the same relations for transitions from the $k$-multiplet
P-wave states to the ground state with the exchange
$\mu^{(-)}\leftrightarrow \mu^{(+)}$.                                  

The first and important prediction is that, assuming one-body 
interactions, S-wave to S-wave electromagnetic transitions are
determined in terms of a {\it single} coupling constant, namely $g_{M1}$.
Using the appropriate charges and masses of the light quarks in Eq.
(\ref{delta+-}), the
radiative coupling $g_2$ is expected to be suppressed by a factor of
{\it three} with respect to the coupling $g_1$. 
One has a total number of {\it six} independent couplings, $f_{E1}^{(K)}$, 
$g_{M2}^{(K)}$ and $f_{E3}^{(K)}$ for transitions from the symmetric P-wave 
to S-wave states and $f_{E1}^{(k)}$, $g_{M2}^{(k)}$ and $f_{E3}^{(k)}$ for 
antisymmetric P-wave down to the ground state decay modes. 
P-wave to P-wave (K-multiplet) radiative channels are determined in terms of 
{\it two} coupling constants which are $G_{M1}^{(K)}$ and $F_{E2}^{(K)}$.
Other two couplings $G_{M1}^{(k)}$ and $F_{E2}^{(k)}$ are also
required for transitions from the k-multiplet down to S-wave states.
Since quadrapole transitions are
expected to dominate the decay rates, with respect to the higher
multipoles, one ends up with only three couplings to describe S- and P-wave
heavy baryon transitions.       

The vanishing of the couplings $f_1^{(K)}$, $F_1^{(K)}$ and $G_1^{(K)}$,
means that, to leading order in HQS
and assuming one-body transition operators,
$\Lambda_{Q1}\rightarrow\Lambda_Q$, 
$\Sigma_{Q1}\rightarrow\Lambda_Q$, 
$\Lambda_{Q1}\rightarrow\Lambda_{Q1}$  
and 
$\Sigma_{Q1}\rightarrow\Sigma_{Q1}$  
radiative decays are all forbidden. Physically, this means that the spins
of the
constituent light quarks are oriented such that a complete cancellation 
arises when both initial and final states are antisymmetric.
This result is in partial agreement with predictions obtained within the 
bound state picture \cite{chow}. This model 
shows that $\Lambda_{c1}\rightarrow\Lambda_c$ radiative decays are
suppressed while the corresponding
bottom transitions are not. On the other hand 
the decay modes $\Sigma_c^{+}\rightarrow\Lambda_c^{+}\;\gamma$ seems not
to be suppressed, which is 
in agreement with predictions use the HHCPT \cite{cho,zd}.  
%=====================================================
\section{Decay rates predictions}
The heavy baryon radiative decay rates are
expected to be saturated by contributions from transitions with the
lowest multipolarity, namely $M1$ for the ground state and $E1$ for the
orbitally excited states. 
Therefore, the determination of the couplings
$g_{M1}$, $f_{E1}^{(K)}$ and $G_{M1}^{(K)}$ is crucial for the
predictions of the electromagnetic decay rates of charmed and bottom
baryons. With the limitations or absence of experimental data, in
particular for S-wave transitions, one thus has to rely on 
specific theoretical models to determine these couplings, a task which
will be postponed for future work. However, results of some  
phenomenological models can still be used 
to fix some of these couplings. 

From light-cone QCD sum rules model
\cite{zd}, for instance, one predicts \be\label{QCD-sum} 
\Gamma_{\Sigma_b\rightarrow\Lambda_b\gamma}=0.29\times
10^{-7}{\mid \vec{q}\mid}^3 {\;\rm MeV} 
\ee
 Using this result and the decay rate formula
\be
\Gamma_{\Sigma_Q\rightarrow\Lambda_Q\gamma}=\frac{4}{3\pi}{\mid 
g_{M1}\mid}^2{\mid
\vec{q}\mid}^3\frac{m_{\Lambda_b}}{m_{\Sigma_b}}, 
\ee 
one obtains
\be\label{gM1}
g_{M1}=2.26\times 10^{-4}\sqrt{\frac{m_{\Sigma_b}}{m_{\Lambda_b}} } 
\;{\rm MeV^{-1}}. 
\ee
Taking the numerical values $m_{\Lambda_b}=5624 {\;\rm MeV}$ and 
$m_{\Sigma_b}=m_{\Lambda_b}+173 {\;\rm MeV}$, as measured by DELPHI
\cite{delphi}, we can determine the
coupling $g_{M1}=0.27\times 10^{-3} {\;\rm MeV}^{-1}$. The
$g_{\Sigma_Q\Lambda_Q\pi}$ coupling was calculated in \cite{pion} using a 
Light-front quark model. Taking this value for the strong coupling,
a simple calculation shows that the branching 
fraction of this decay mode relative to that of the single pion
transition is about $0.25\%$. The decay rate of the corresponding charmed
baryons, on the
other hand, are predicted to be  
\be
\Gamma_{\Sigma_c\rightarrow\Lambda_c\gamma}=87 {\; \rm KeV}.
\ee
This value is lower than the one calculated in HHCPT \cite{cheng} but
 a bit higher than that reported in \cite{iklr}. Neglecting the E2
transitions, which are suppressed with respect to M1, and using the value
of the coupling $g_{M1}$ given by Eq. (\ref{gM1}) we can estimate the decay
rate 
\be
\Gamma_{\Sigma^{0*}_b\rightarrow\Sigma_b^0\gamma}=1.72 {\; \rm KeV},
\ee
which agrees with light-cone QCD sum rules results \cite{zd}. One
also can calculate the charmed baryon decay rate
\be
\Gamma_{\Sigma^{+*}_c\rightarrow\Sigma_c^+\gamma}=0.19 {\; \rm KeV}.
\ee 
We notice that this rate, which is very close to the one reported in
\cite{iklr}, idicates that charmed baryon modes are suppressed by a factor
of 10 with respect to the bottom ones.    

The coupling $f_{E1}^{(K)}$ can be fixed from HHCPT predictions 
\cite{chow}, which gives 
\be
\Gamma_{\Lambda_{Q1}\rightarrow\Sigma_Q\gamma}=0.136 \; {\mid c_{RS}\mid}^2
{\; \rm
MeV},
\ee
here $c_{RS}$ is the HHCPT coupling $g_{\Sigma\Lambda_{c1}\gamma}$. 
On the basis of dimensional analysis this coupling is
expected to be of order unity, which is also supported by dynamical
calculations \cite{iklr}. Now using the decay rate formula 
\be
\Gamma_{\Lambda_{Q1}\rightarrow\Sigma_Q\gamma}=\frac{4}{9\pi}{\mid  
f_{E1}^{(K)}\mid}^2{\mid
\vec{q}\mid}^3\frac{m_{\Lambda_{c1}}}{m_{\Sigma_c}},   
\ee
one gets 
\be
f_{E1}^{(K)}=0.46\times 10^{-3} {\; \rm MeV}^{-1}. 
\ee
If we take $m_{\Lambda_{b1}}=5900 {\;\rm MeV}$ and $m_{\Lambda_{b2}}=5926
{\;\rm MeV}$, as predicted in large-$N_c$ limit \cite{cw}, we
find
\be
\Gamma_{\Lambda_{b1}\rightarrow\Sigma_b\gamma}=0.032 {\; \rm
MeV} {\; \rm and \; } \Gamma_{\Lambda_{b2}\rightarrow\Sigma_b\gamma}=0.063 
{\; \rm MeV}.
\ee
These prediction are at least a factor of two lower than the results
obtained using a bound state picture \cite{chow}. For charmed baryons we 
predict 
\be
\Gamma_{\Lambda_{c1}(2593)\rightarrow\Sigma_c\gamma}=0.071 {\; \rm
MeV} {\; \rm and \; }
\Gamma_{\Lambda_{c1}(2593)\rightarrow\Sigma^{*}_c\gamma}=0.011
{\; \rm MeV},
\ee
and 
\be
\Gamma_{\Lambda^{*}_{c1}(2626)\rightarrow\Sigma_c\gamma}=0.13 {\; \rm
MeV} {\; \rm and \; }
\Gamma_{\Lambda^{*}_{c1}(2626)\rightarrow\Sigma^{*}_c\gamma}=0.032
{\; \rm MeV},
\ee
which are close to the RTQM results \cite{iklr} but at least a factor of
three larger than the bound state predictions \cite{chow}.

From Eq. (\ref{s-s}), with $m_u=m_d$ and using the appropriate
values for the light quark
charges in Eq. (\ref{delta+-}), we also predict the electromagnetic decay
rates ratios $\Gamma_{\Sigma_c^{++}\rightarrow\Sigma_c^{++}}:
\Gamma_{\Sigma_c^{0}\rightarrow\Sigma_c^{0}}:
\Gamma_{\Sigma_c^{+}\rightarrow\Sigma_c^{+}}$ to be $16:4:1$. 
The same ratios are also obtained for
$\Gamma_{\Sigma_b^{+}\rightarrow\Sigma_b^{+}}:
\Gamma_{\Sigma_b^{-}\rightarrow\Sigma_b^{-}}:
\Gamma_{\Sigma_b^{0}\rightarrow\Sigma_b^{0}}$ which are in very good
agreement with the ratios $2.2:0.54:0.14$ obtained in the Light-cone QCD
sum rules \cite{zd}. Furthermore, 
if one neglects the magnetic quadrapole transition, which is suppressed
relative to the electric dipole contributions, 
the same ratios can be obtained for the corresponding decay rates of 
$\Sigma_{Q1}\rightarrow\Sigma_Q\;\gamma$,
$\Sigma_{Q2}\rightarrow\Sigma_Q\;\gamma$,   
$\Lambda_{Q1}\rightarrow\Sigma_Q\;\gamma$ and
$\Lambda_{Q2}\rightarrow\Sigma_Q\;\gamma$. 
%==========================================================  
\section{Conclusions}
Radiative decays of heavy baryons have been investigated to leading order
in HQS supplemented by of the $SU(2N_f)\times O(3)$ diquark
symmetries.
Neglecting quadrapole transitions, which are suppressed in the
$m_Q\rightarrow\infty$ limit, only {\bf three} couplings are needed
to describe electromagnetic transitions of ground state and the lowest
orbitally excited heavy baryon states. The decay rates and branching
ratios of some radiative transitions for heavy
baryons are predicted. Our results are in agreement with most of the
other theoretical model calculations for both charmed and bottom baryons.
We notice that simple quark model calculations and the use of 
other theoretical models, however, shows that many of the charmed  
baryon decay rates are suppressed with respect to the bottom once.
 All the allowed photon transitions among heavy baryon states can be
predicted as soon as some more
experimental data become available to fix these few effective
couplings. 
           
To conclude we would like to stress that, these predictions are
leading order results and contributions from two-body and $\frac{1}{m_Q}$
corrections might be important, especially for charmed baryons. On the
other hand our predictions for bottom baryons will receive a maximum of up
to $10\%$ contributions from the non-leading order. Since the CLEO, E781 
experiments and some other facilities are expected to collect more data in
the next few years, we are confident that they can measure some of the
heavy baryon electromagnetic channels to confirm our constituent quark
model predictions. 

\setcounter{equation}{0}
\def\theequation{A.\arabic{equation}}
\section*{Appendix \\ P-wave to P-wave Electromagnetic transitions} 
 In this appendix, we shall follow section 2 and 3 to analyze P-wave to
P-wave single-photon transitions. Neglecting transitions higher than
quadrapole, which are small, HQS predicts that these transitions 
are determined in terms of fifteen independent couplings for each of the
K- and k-multiplets. The appropriate and covariant decay matrix elements
are summarized in Table (\ref{tab-HQS-b}). Employing the light diquark
symmetry one can relate these couplings to only two constituent quark
model couplings $G_{M1}$ and $F_{E2}$. The one-body effective
operators mediating these transitions
among K-excited P-wave states are given by 
\be\label{M1-K(p-p)} 
\left({\cal O}_{\lambda\zeta}^{M1}\right)^{\alpha\beta}_{\alpha^{'}
\beta^{'}} =i\frac{G_{M1}^{(K)}}{2} 
\left(\mu_{q_1}(\sigma^{\delta\zeta})^{\alpha}_{\alpha^{'}}
 \otimes (\eins)^{\beta}_{\beta^{'}}
        + \,\mu_{q_2}(\eins )^{\alpha}_{\alpha^{'}}\otimes 
(\sigma^{\delta\zeta})^{\beta}_{\beta^{'}}\right)\, F_{\delta\lambda},
\ee                
and 
\[
\left({\cal O}_{\lambda\zeta}^{E2}\right)^{\alpha\beta}_{\alpha^{'}
\beta^{'}} =i\frac{F_{E2}^{(K)}}{2} 
\left(\mu_{q_1}(\sigma^{\delta\zeta})^{\alpha}_{\alpha^{'}}
 \otimes (\eins)^{\beta}_{\beta^{'}}
       + \,\mu_{q_2}(\eins )^{\alpha}_{\alpha^{'}}\otimes 
(\sigma^{\delta\zeta})^{\beta}_{\beta^{'}}\right) \] \nonumber  
\be\label{E2-K(p-p)}  
\,\left[q_{\lambda}F_{\eta\delta}v^{\eta}+(v\cdot q)\; 
F_{\delta\lambda}\right].
\ee 
Similar operators can be written for transitions among k-multiplet states
with new coupling $G_{M1}^{(k)}$ and $F_{E2}^{(k)}$. 
It is important to mention that the effective operators given by 
Eq. (\ref{M1s-s}) and Eqs. (\ref{E1-K(p-s)}-\ref{E3-K(p-s)}) and
(\ref{M1-K(p-p)}-\ref{E2-K(p-p)}) are unique. 
This can be proved with the help of the relation \cite{man} 
\be 
\frac{1}{2}(\vslash+1)\sigma^{\mu\nu}\frac{1}{2}(\vslash+1)=
v_{\alpha}\epsilon^{\alpha\mu\nu\beta}s_{\beta},
\ee
where $s_{\beta}$ are the three spin vectors. The absence of
experimental data as well as predictions using theoretical models makes it
impossible, for the time being, to fix the two couplings $G_{M1}^{(k)}$
and $F_{E2}^{(k)}$.
Therefore, we will not present any phenomenological values regarding the
 decay rates among P-wave states.    
%****************** T A B L E  3 *****************************************
\begin{table}
\caption[dumy16]{\label{tab-HQS-b} Light-diquark matrix elements for
P-wave to P-wave K-multiplet transitions, the heavy quark matrix elements
${\cal H}^{\mu_1 \cdots\mu_{j_1}}_{\nu_1\cdots \nu_{j_2}}(v) $ 
are similar to those in Table (\ref{tab-HQS-b}).  
Here, we have neglected transitions with $J_{\gamma}>2$.  }
%\begin{center}   
\vspace{3mm}
\renewcommand{\baselinestretch}{1.2}
\small \normalsize
\begin{center}
%\footnotesize {
%\scriptsize 
\small{    
\begin{tabular}{|c||c|c|}
\hline \hline
 $B_{Qi} \rightarrow B_{Qi}^{'}\gamma$ &  Multipole state  &
${\cal M}^{\nu_1 \cdots\nu_{j_2}}_{\mu_1\cdots \mu_{j_1}}(v,q)$ \\ 
\hline \hline  
$\Lambda_{Q1}\rightarrow \Lambda_{Q1}$ & $ M1,E2$ &
$ \begin{array}{c} 
G_1^{(K)} F_{\alpha \beta} g_{\mu}^{\alpha} g_{\nu}^{\beta}+  \\ 
  F_1^{(K)} F_{\alpha \beta}(2q_{\mu}
       g_{\nu}^{\alpha}v^{\beta}+v \hspace{-0.7mm}\cdot\hspace{-0.7mm}q
       g_{\mu}^{\alpha} g_{\nu}^{\beta})               
 \end{array}                          
 $\\                      
\hline
$\Sigma_{Q0}\rightarrow \Lambda_{Q1}$ & $ M1$ &
$iG_2^{(K)}{\tilde F}^{\alpha \beta} g_{\mu\alpha} v_{\beta}         
 $\\              
\hline          
$\Sigma_{Q1}\rightarrow \Lambda_{Q1}$ & $ M1,E2$ &
$ \begin{array}{c} 
G_3^{(K)} F_{\alpha \beta} g_{\mu}^{\alpha} g_{\nu}^{\beta}+  \\ 
  F_3^{(K)} F_{\alpha \beta}(2q_{\mu}
       g_{\nu}^{\alpha}v^{\beta}+v \hspace{-0.7mm}\cdot\hspace{-0.7mm}q
       g_{\mu}^{\alpha} g_{\nu}^{\beta})               
 \end{array}                          
 $\\ 
\hline                    
$\Sigma_{Q1}\rightarrow \Sigma_{Q0}$ & $ M1$ &  
 $ i G_4^{(K)}{\tilde F}^{\alpha \beta}
       g_{\mu\alpha} v_{\beta}  $\\ 
\hline
$\Sigma_{Q1}\rightarrow \Sigma_{Q1}$ & $ M1,E2$ &
 $  \begin{array}{c} 
G_5^{(K)} F_{\alpha \beta} g_{\mu}^{\alpha} g_{\nu}^{\beta}+  \\ 
  F_5^{(K)} F_{\alpha \beta}(2q_{\mu}
       g_{\nu}^{\alpha}v^{\beta}+v \hspace{-0.7mm}\cdot\hspace{-0.7mm}q
       g_{\mu}^{\alpha} g_{\nu}^{\beta})               
 \end{array}                          
 $\\                       
\hline           
$\Sigma_{Q2}\rightarrow \Lambda_{Q1}$ & $M1,E2$ &  
$ \begin{array}{c}  
iG_6^{(K)} {\tilde F}^{\alpha \beta}g_{\mu_{1}\alpha}
g_{\mu_{2}\nu}v_{\beta}+\\
 iF_6^{(K)}{\tilde F}^{\alpha \beta} 
 ( 2q_{\mu_{1}} g_{\mu_{2}\alpha}g_{\nu\beta} + \\
v \! \cdot \! q g_{\mu_{1} \nu} g_{\mu_{2}\alpha}v_{\beta} )
\end{array} 
$\\       
\hline  
%\hspace*{-0.8cm} 
$\Sigma_{Q2}\rightarrow \Sigma_{Q0}$ & $E2$ &  
 $ F_7^{(K)}F_{\alpha \beta} q_{\mu_{1}}  
  g_{\mu_{2}}^{\alpha}v^{\beta}
$\\       
\hline               
$\Sigma_{Q2}\rightarrow \Sigma_{Q1}$ & $M1,E2$ &  
$  \begin{array}{c}  
i G_8^{(K)} {\tilde F}^{\alpha \beta}g_{\mu_{1}\alpha}
g_{\mu_{2}\nu}v_{\beta}+\\
 i F_8^{(K)}{\tilde F}^{\alpha \beta}  
 ( 2q_{\mu_{2}} g_{\mu_{1}\alpha}g_{\nu\beta} +\\
v \! \cdot \! q g_{\mu_{2} \nu} g_{\mu_{1}\alpha}v_{\beta} )
\end{array} $\\                 
\hline 
$\Sigma_{Q2}\rightarrow \Sigma_{Q2}$ & $M1,E2$ &  
$  \begin{array}{c}  
G_9^{(K)}  F_{\alpha \beta}g_{\mu_{1}}^{\alpha} g_{\nu_{1}}^{\beta}
g_{\mu_{2}\nu_2}+ \\
 F_9^{(K)} 
F_{\alpha \beta}
 ( 2q_{\mu_{1}} g_{\nu_{1}}^{\alpha}g_{\mu_2\nu_2}v^{\beta} + \\
v \! \cdot \! q g_{\mu_{1}}^{\alpha}g_{\nu_{1}}^{\beta} g_{\mu_2\nu_2} )
\end{array} $\\                                                                           
\hline \hline              
\end{tabular}  
}
\end{center}
\renewcommand{\baselinestretch}{1}
\small \normalsize
\end{table}      
%\newpage
%----------------------------------------------------------------------

\end{document}